\documentstyle[epsfig,12pt]{article}

\newcommand{\bce}{\begin{center}} 
\newcommand{\ece}{\end{center}}
\newcommand{\beq}{\begin{equation}}
\newcommand{\eeq}{\end{equation}}
\newcommand{\bea}{\vspace{0.25cm}\begin{eqnarray}}
\newcommand{\eea}{\end{eqnarray}}

\newcommand{\ba}{\begin{array}}
\newcommand{\ea}{\end{array}}

\newcommand{\rhs}{{\sl rhs~}}

\newcommand{\doublespace}{
    \renewcommand{\baselinestretch}{1.6}\large\normalsize}

\def\lsim{\mathrel{\rlap{\lower4pt\hbox{\hskip1pt$\sim$}}
    \raise1pt\hbox{$<$}}}         
\def\gsim{\mathrel{\rlap{\lower4pt\hbox{\hskip1pt$\sim$}}
    \raise1pt\hbox{$>$}}}         

\def\beq{\begin{equation}}
\def\endeq{\end{equation}}
\def\arr{\begin{eqnarray}}
\def\endarr{\end{eqnarray}}
\makeindex

  \advance\oddsidemargin  by -1in
  \advance\evensidemargin by -1in
  \advance\topmargin      by -0.5in
\begin{document}

\vspace{2.0cm}

\begin{flushright}
ITEP(Ph)-2005-07
\end{flushright}

\vspace{1.0cm}

\begin{center}
{\Large \bf 
Left-right asymmetry of nuclear shadowing \\
 in charged current DIS}

\vspace{1.0cm}

{\large\bf R.~Fiore$^{1}$  and 
V.R.~Zoller$^{2}$}

\vspace{1.0cm}

$^1${\it Dipartimento di Fisica, 
Universit\`a della Calabria\\
and\\
Istituto Nazionale di Fisica Nucleare,\\
Gruppo Collegato di Cosenza,\\
I-87036 Rende, Cosenza, Italy}\\
fiore@cs.infn.it\\ 
$^2${\it
ITEP, Moscow 117218, Russia}\\
zoller@itep.ru
\vspace{1.0cm}\\
{ \bf Abstract }\\
\end{center}
We study the shadowing effect in  highly asymmetric
diffractive interactions of left-handed  and right-handed 
$W$-bosons with atomic nuclei. 
The target nucleus is found to be quite  transparent for
the charmed-strange  Fock component  of the light-cone
$W^+$  in the helicity state $\lambda=+1$
and  rather opaque for the $c\bar s$ dipole  with  $\lambda=-1$.
  The shadowing  correction to the structure function
$\Delta xF_3 =
xF_3^{\nu N}-xF_3^{\bar\nu N}$ extracted  from $\nu Fe$ and $\bar\nu Fe$ data
is shown to make up  about $20\%$ in the kinematical range of CCFR/NuTeV.

\doublespace
\pagebreak


\section{Introduction}
There are several well established facts about  nuclear shadowing
in photo- and electro-production,
the phenomenon expected since long ago \cite{STOD, GRIB, KANCH, NNNVIZ}:
\begin{itemize}
\item
{the phenomenon does exist \cite{EMC},}
\item
{ it scales \cite{NZ91}},
\item{ it is dominated by large, hadronic size  color $q\bar q$-dipoles,
$r\sim {m^{-1}_q},$ that the virtual photon transforms into at a
large distance upstream the target \cite{NZ91}.}
\end{itemize}
The latter point implies that shadowing in photo- and electro-production
is $\propto m_q^{-2}$ and is rather small in the charm structure function of
nuclei.

The situation is quite different in the  charged current (CC) neutrino
deep-inelastic scattering (DIS). At small values of Bjorken $x$ the
charm production in CC  DIS is driven by the $W^+$-gluon  fusion,
\beq
W^+g\to c\bar s.
\label{eq:WG}
\eeq
In the process (\ref{eq:WG}) charm is inseparable from strangeness
and the size of the relativistic  heavy-light
color dipole $c\bar s$  depends strongly on the  momentum partition
in the light-cone  $W^+$-decay. The latter is determined  by the light-cone
wave function (LCWF) of the Fock state  $|c\bar s\rangle$  and depends  on
the helicity of $W^+$-boson \cite{FZ}.
In this communication  we demonstrate that for the
left-handed $W^+$ shadowing is $\propto 1/\mu^2$ and for the  right-handed
$W^+$ shadowing is $\propto 1/m^2$, where $\mu$ and $m$ are the strange
and charmed  quark masses, respectively. Therefore, in spite of the
presence of heavy quark in the color singlet  state propagating through
the target nucleus, one can expect
considerable shadowing corrections to the nucleon  structure function
 $$xF_3\propto \sigma_L-\sigma_R$$
extracted from nuclear data. Here $\sigma_L$ and $\sigma_R$ are the
absorption cross sections
for the left-handed and right-handed $W^+$-bosons, respectively.
This structure function is known to be
determined by the process (\ref{eq:WG}) if $x$ is small enough.
The latter  is associated with the t-channel vacuum exchange or
the sea-quark  contribution to  $xF_3$.
However, the data available were taken at  moderately small-$x$ and
  the valence term  in  $xF_3$ cannot be neglected \cite{FZ}.
Free of the non-vacuum contributions is
the combination of $\nu N$ and $\bar\nu N$
structure functions
\beq
\Delta xF_3=xF_3^{\nu N}-xF_3^{\bar\nu N}.
\label{eq:DXF3}
\eeq
Hereafter, $N$ stands for  an iso-scalar nucleon.
Our finding is that the  shadowing  correction to
 $\Delta xF_3$ extracted from  $\nu Fe$ and $\bar\nu Fe$
data amounts to $20-25\%$ in the  kinematical range of
 CCFR/NuTeV experiment
\cite{CCFR, CCFR2, CCFR3}.

Different approaches to nuclear shadowing in the neutrino DIS
have been discussed previously (see \cite{Bell, More}). 
In this communication
 we develop  the color dipole 
description of the phenomenon with particular emphasis on the 
left-right asymmetry effect specific for the charged current DIS.

\section{Color dipole description of CC DIS off nuclei}

 The interaction of high-energy neutrino
with the target nucleus can be  treated as mediated by the interaction of
the quark-antiquark  dipole that 
the virtual $W^+$ transforms into. At small values of Bjorken $x$
this transition takes place at a large  distance $l$, upstream the target:
\beq
l\sim {1\over k_L}\sim {1\over xm_N} .
\label{eq:L}
\eeq
Here $k_L$ is the longitudinal momentum transfer
in the transition  (\ref{eq:WG}), $x=Q^2/2pq$, $p$ and $q$ are the 
target nucleon and the virtual $W^+$
four-momenta, respectively,  
$q=(\nu,0,0,\sqrt{\nu^2+Q^2})$, $Q^2=-q^2$ and $2pq=2m_N\nu$.

Interaction of the color dipole  of size ${\bf r}$  
 with the target nucleon is described by the beam- 
 and flavor-independent color dipole cross section 
$\sigma(x,r)$ \cite{NZ91, NZ94, M}.  At small $x$ the dipole size 
${\bf r}$ is a conserved quantum number. Therefore,
the contribution of the excitation of open 
charm/strangeness to the nuclear  absorption cross section for 
left-handed, $(\lambda=L=-1)$ and  right-handed, $(\lambda=R=+1)$,   
$W^+$-boson of virtuality $Q^2$ 
is given by the color dipole
factorization formula \cite{ZKL,BBGG}
\bea
\sigma^A_{\lambda}(x,Q^{2})
=\langle \Psi_{\lambda}| \sigma^A(x,r)|\Psi_{\lambda}\rangle\nonumber\\
=\int dz d^{2}{\bf{r}} \sum_{\lambda_1,\lambda_2}
|\Psi_{\lambda}^{\lambda_1,\lambda_2}(z,{\bf{r}})|^{2} 
\sigma^A(x,r)\,,
\label{eq:FACTOR}
\eea
where \cite{GLAUBER}
\beq
\sigma^A(x,r)=
2\int d^{2}{\bf{b}}
\left\{1-\exp\left[-{1\over 2}\sigma(x,r)T(b)\right]\right\}. 
\label{eq:SIGA}
\eeq
Here $T(b)$ is the the optical  thickness of a nucleus,  
\beq 
T(b)=\int_{-\infty}^{+\infty} dz n(\sqrt{z^2+b^2}),
\label{eq:TB}
\eeq 
${\bf b}$ is the impact parameter and $n(r)$ is the nuclear matter density 
normalized as follows: 
\beq
\int d^3r n(r)=A.
\label{eq:NA}
\eeq 
It is assumed that $A\gg 1$. 
One can expand the exponential in Eq.~(\ref{eq:SIGA}) to  separate  
the impulse approximation term 
and the  shadowing correction, $\delta\sigma^A_{\lambda}$, 
in  (\ref{eq:FACTOR}),
\beq
\sigma^A_{\lambda}=
A\sigma_{\lambda}
-\delta\sigma^A_{\lambda}.
\label{eq:NUCLEAR}
\eeq
Here
\bea
\sigma_{\lambda}=
\langle\Psi_{\lambda}|\sigma(x,r)  |\Psi_{\lambda} \rangle\nonumber\\
=\int dz d^{2}{\bf{r}} \sum_{\lambda_1,\lambda_2}
|\Psi_{\lambda}^{\lambda_1,\lambda_2}(z,{\bf{r}})|^{2} 
\sigma(x,r)\,.
\label{eq:SIGAV}
\eea
To the lowest order  in $\sigma T$ the shadowing term  reads 
\beq
\delta\sigma^A_{\lambda}\simeq
{\pi\over 4} \langle \sigma_{\lambda}^2\rangle
{\cal S}^2_A(k_L)\int db^2 T^2(b),
\label{eq:SHADOW}
\eeq
where
\bea
\langle \sigma_{\lambda}^2\rangle=
\langle\Psi_{\lambda}|\sigma(x,r)^2|\Psi_{\lambda} \rangle\nonumber\\
=\int dz d^{2}{\bf{r}} \sum_{\lambda_1,\lambda_2}
|\Psi_{\lambda}^{\lambda_1,\lambda_2}(z,{\bf{r}})|^{2} 
\sigma^2(x,r).
\label{eq:SIG2AV}
\eea
The longitudinal nuclear form factor ${\cal S}_A(k_L)$ in
Eq.~(\ref{eq:SHADOW}) takes care about the coherency constraint,
\beq
l\gg R_A.
\label{eq:COHCON}
\eeq
The approximation (\ref{eq:SHADOW}) represents the driving term of shadowing, 
the double-scattering term. It is reduced by the higher-order rescatterings
by about $30\%$ for iron  and $50\%$ for lead nuclei. This 
 accuracy is  quite sufficient   for  order-of-magnitude  estimates. 
The numerical calculations presented below are done for the full Glauber series
(\ref{eq:SIGA}),  
\beq
\delta\sigma^A_{\lambda}=\pi{\cal S}^2_A(k_L)\sum_{n=2}^{\infty}
{(-1)^n\langle \sigma_{\lambda}^n\rangle\over n! 2^{n-1}}
\int db^2 T^n(b)\,,
\label{eq:GSERIES}
\eeq 
where the effect of finite coherence length is modeled by the factor
${\cal S}^2_A(k_L)$ in  $\rhs$. A consistent description of the latter
effect in electro-production was obtained in Ref.~\cite{BGZ98} based on  the 
light-cone path integral technique of  Ref.~\cite{BGZ96}.
For related numerical studies of nuclear shadowing in electro-production
see \cite{KRT}.

The LCWF, $\Psi_{\lambda}^{\lambda_1,\lambda_2}(z,{\bf{r}})$,
in Eqs.~(\ref{eq:FACTOR}), (\ref{eq:SIGAV}) and (\ref{eq:SIG2AV}) describes
the  Fock  state $|c\bar s\rangle$ with the $c$-quark
carrying the fraction $z$ of the $W^+$ light-cone momentum and the
$\bar s$-quark with momentum fraction $1-z$. The $c$- and $\bar s$-quark
helicities are  $\lambda_1=\pm 1/2$ and  $\lambda_2=\pm 1/2$, respectively.
This wave function  derived in Ref.~\cite{FZ} is  found in Appendix.

The diagonal elements of  the density matrix 
\beq
\rho_{\lambda\lambda^{\prime}}
=\sum_{\lambda_1,\lambda_2}\Psi_{\lambda}^{\lambda_1,\lambda_2}
\left(\Psi_{\lambda^{\prime}}^{\lambda_1,\lambda_2}\right)^*
\label{eq:RHO}
\eeq
for $\lambda=\lambda^{\prime}=L,R$ 
entering Eqs.~ 
(\ref{eq:FACTOR}), (\ref{eq:SIGAV}) and (\ref{eq:SIG2AV}) are as follows: 
\bea
\rho_{RR}(z,{\bf r})=\left|\Psi_{R}^{+1/2, +1/2}\right|^2
+\left|\Psi_{R}^{-1/2, +1/2}\right|^2\nonumber\\
={{8\alpha_W N_c}\over (2\pi)^2}(1-z)^2\left[m^2 K^2_0(\varepsilon r)
+\varepsilon^2 K^2_1(\varepsilon r)\right]
\label{eq:RHOR}
\eea
and
\bea
\rho_{LL}(z,{\bf r})=\left|\Psi_{L}^{-1/2,-1/2}\right|^2
+\left|\Psi_{L}^{-1/2,+1/2}\right|^2\nonumber\\
={{8\alpha_W N_c}\over (2\pi)^2}z^2\left[\mu^2 K^2_0(\varepsilon r)
+\varepsilon^2 K^2_1(\varepsilon r)\right],
\label{eq:RHOL}
\eea
where
\beq
\varepsilon^2=z(1-z)Q^2+(1-z)m^2+z\mu^2
\label{eq:VAREP}
\eeq
 and   $K_{\nu}(x)$ is the modified Bessel function. 
The $c$-quark 
and $\bar s$-quark masses are $m$ and $\mu$, respectively.

Consequences of the  striking momentum partition asymmetry of both  
$\rho_{LL}$ and $\rho_{RR}$ were studied  in Ref.~\cite{FZ}.
This asymmetry was found to determine the 
vacuum exchange contribution to the 
structure function $xF_3$ of CC DIS: 
\beq
2xF_3(x,Q^2)={Q^2\over 4\pi^2\alpha_{W}}
\left[\sigma_{L}(x,Q^{2})-\sigma_{R}(x,Q^{2})\right].
\label{eq:F3}
\eeq

\section{Left and right: different scales - different dynamics}

The CCFR/NuTeV structure functions 
$xF^{\nu N}_3$ and $xF^{\bar\nu N}_3$ are  extracted from the 
$\nu Fe$  and $\bar\nu Fe$ data \cite{CCFR,CCFR2,CCFR3}. 
 To estimate  the strength of the nuclear shadowing effect in
$xF_3$ at high $Q^2$ such that 
\beq
{m^2\over Q^2}\ll 1,\,\,{\mu^2\over Q^2}\ll 1  
\label{eq:DLLA}
\eeq
one should take into account that the dipole cross section $\sigma(x,r)$ 
in Eqs.~(\ref{eq:SHADOW}) and (\ref{eq:SIG2AV}) is related to the un-integrated
gluon structure function
${{\cal F}(x,\kappa^2)}={\partial G(x,\kappa^2)/\partial\log{\kappa^2}},$
 as follows \cite{FACTOR}: 
\bea
\sigma(x,r)={\pi^2 \over N_c}r^2\alpha_S(r^2)
\int{d\kappa^2\kappa^2\over (\kappa^2+\mu_G^2)^2}
{4[1-J_0(\kappa r)]\over \kappa^2r^2}{{\cal F}(x_g,\kappa^2)}.
\label{eq:SIGMA}
\eea
In the  Double Leading Logarithm Approximation (DLLA),
i.e. for small dipoles, we have
\bea
\sigma(x,r)\approx {\pi^2\over N_c} r^2\alpha_S(r^2)
G(x_g,C/r^2),
\label{eq:SMALL}
\eea
where $\mu_G=1/R_c$ is the inverse correlation radius of perturbative gluons
and $C\simeq 10$ comes from properties of the Bessel function $J_0(y)$.
Because of the scaling violation $G(x,Q^2)$  rises with $Q^2$, but the product
$\alpha_S(r^2)G(x,C/r^2)$ is approximately flat in $r^2$. 
Let us estimate first the
contribution to $\langle\sigma^2_{\lambda}\rangle$ coming from
the P-wave term, $\varepsilon^2 K_1(\varepsilon r)^2$, in Eqs.~(\ref{eq:RHOR})
and (\ref{eq:RHOL}). 
 The asymptotic behavior of the
Bessel function, $K_1(x)\simeq \exp(-x)/\sqrt{2\pi/x}$ makes the 
$\bf{r} $-integration rapidly convergent at $\varepsilon r > 1$.
 Integration over $\bf{r}$ in Eq.~(\ref{eq:SIG2AV}) yields
\beq
\langle\sigma^2_{L}\rangle\propto \int_0^1 dz {z^2\over \varepsilon^4}\propto 
{1\over Q^2 \mu^2}
\label{eq:SIGL}
\eeq
and similarly 
\beq
\langle\sigma^2_{R}\rangle\propto \int_0^1 dz {(1-z)^2\over \varepsilon^4}
\propto  {1\over Q^2m^2}.
\label{eq:SIGR}
\eeq
We are not surprised to see that shadowing is the scaling, rather than the 
higher twist,$1/Q^2$, effect.
Obviously, the integral (\ref{eq:SIGL}) is dominated by $z\gsim 1-\mu^2/Q^2$
i.e., by $\varepsilon^2\sim \mu^2$ and, consequently, by 
$r^2\sim 1/\varepsilon^2\sim  1/\mu^2$. A comparable  contribution
to the integral (\ref{eq:SIGL}) comes from the S-wave term
$\propto \mu^2 K_0(\varepsilon r)^2$ in $\rho_{LL}$. In Eq.~(\ref{eq:SIGR})
the integral
is dominated by  $z\lsim m^2/Q^2$, corresponding to 
$\varepsilon^2\sim m^2$. Therefore, 
$r^2\sim 1/\varepsilon^2\sim  1/m^2$.
Thus, we conclude that the 
typical dipole sizes which dominate $\sigma_{\lambda}$ and
$\langle \sigma_{\lambda}^2\rangle$ are very different.
In Ref.~\cite{FZ} basing on the color dipole approach  we found 
the scaling cross sections 
$\sigma_L$ and $\sigma_R$ $\propto 1/Q^2$
times the Leading-Log scaling violation factors
$\propto\log Q^2/\mu^2$ and $\propto\log Q^2/m^2$, respectively. 
The scaling violations  were found to be  (logarithmically) dominated by 
\beq 
r^2\sim 1/Q^2.
\label{eq:1Q2}
\eeq
On the contrary, the contribution of small-size  dipoles,
$\sim 1/Q^2$, to $\langle \sigma_{\lambda}^2\rangle$,  defined in
Eq.~(\ref{eq:SIG2AV}), proved to be  negligible. 
At $\lambda=-1$ $\langle \sigma_{\lambda}^2\rangle$
 is dominated by large hadronic size $c\bar s$-dipoles, $ r\sim 1/\mu$. 
Consequently,  
\beq 
\delta\sigma^A_L\propto  1/\mu^2.
\label{eq:S2L}
\eeq
At $\lambda=+1$ a typical $c\bar s$-dipole  is rather  small, $r\sim 1/m$,
and $\delta\sigma^A_R$ is small as well:
\beq 
\delta\sigma^A_R\propto 1/m^2.
\label{eq:S2R}
\eeq
 Thus, there is a sort of filtering phenomenon,
the target nucleus absorbs the $c\bar s$ Fock component of $W^+$
 with $\lambda=-1$, but 
is nearly transparent for $c\bar s$ states with opposite helicity,
  $\lambda=+1$.

In a region of very  small $Q^2$ a 
considerable asymmetry of the shadowing effect 
is expected also because at  $Q^2\to 0$ 
\beq
\langle \sigma^2_{L}\rangle
\propto {1\over m^2\mu^2}\gg
 \langle \sigma^2_{R}\rangle\propto {1\over m^4}.
\label{eq:SMALLQ1}
\eeq 
 However,
at currently available $x\gsim 0.01$  both the mass threshold effect
and nuclear form factor   suppress
$\delta\sigma^A_{L,R}$  at $Q^2\lsim (m+\mu)^2$.

\section{Shadowing in $\Delta x F_3$: 
the magnitude of effect}

\begin{figure}[h]
\psfig{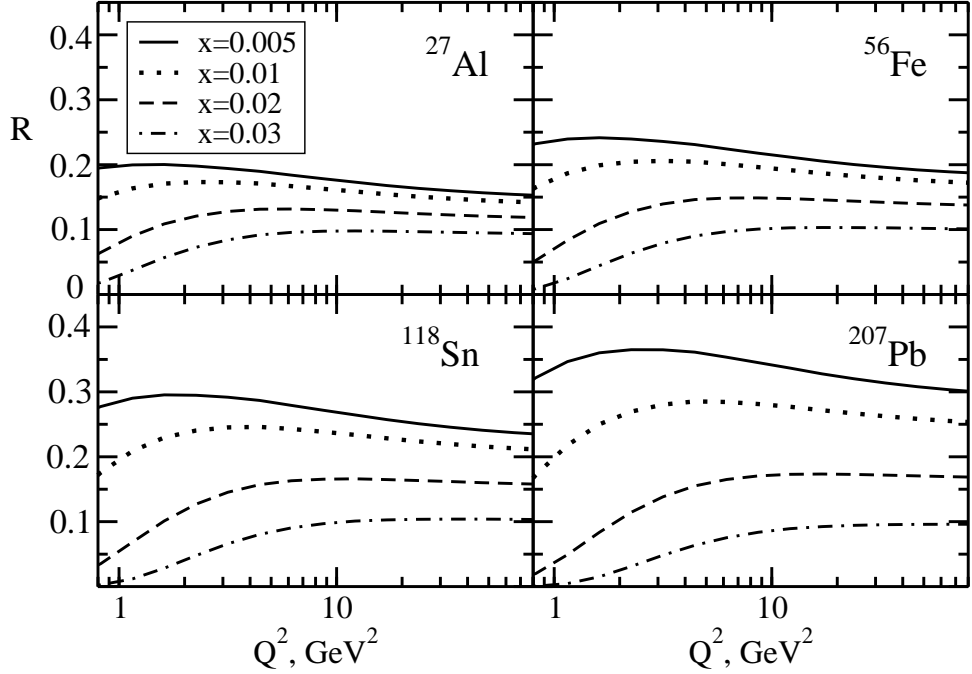}
\vspace{-0.5cm}
\caption{The shadowing ratio $R$ as a function of $Q^2$ for 
several values of $x$
 calculated from the nuclear charge 
densities of Ref.\cite{DEVRIES} for some sample nuclei.}
\label{fig:RNUC}
\end{figure}

It should be repeated
that we  focus on  the vacuum exchange contribution 
to $xF_3$ corresponding
to the excitation of the  $c\bar s$-pair in the process (\ref{eq:WG}).
Therefore, the structure function $xF_3$ 
differs from zero only  due to the strong left-right asymmetry of the 
light-cone  Fock state $|c\bar s\rangle$. This contribution to $xF_3$
 can be reinterpreted
in terms of the sea-quark densities of the target nucleon/nucleus.
At currently available $x\gsim 0.01$, in addition to the sea, 
 the valence contribution  must be taken into account.
 The valence term, $xV$ is the same
for both  $\nu N$ and 
$\bar\nu N$ structure functions of an  iso-scalar nucleon. 
The sea-quark term in $xF^{\nu N}_3$ 
denoted by $xS$ has opposite
sign for $xF^{\bar\nu N}_3$. Indeed,  the substitution 
$m\leftrightarrow\mu$ in Eqs.~(\ref{eq:RHOR}) and (\ref{eq:RHOL}) entails
$\sigma_L\leftrightarrow\sigma_R$.
Hence,
\beq
xF^{\nu N}_3=xV+xS
\label{eq:SVNU}
\eeq
and
\beq
xF^{\bar\nu N}_3=xV-xS.
\label{eq:SVANU}
\eeq
One can combine the $\nu N$ and 
$\bar\nu N$ structure functions  
to isolate the Pomeron exchange term. Indeed, from Eqs.~(\ref{eq:DXF3}), 
(\ref{eq:SVNU}) and (\ref{eq:SVANU}) it follows that
\beq
\Delta x F_3=2xS.
\label{eq:DF3}
\eeq
The extraction of $\Delta x F_3$
from CCFR $\nu_{\mu}Fe$ and $\bar\nu_{\mu}Fe$ differential cross section
in a model-independent way has been reported in Ref.~\cite{CCFR3}.
From Eq.(\ref{eq:NUCLEAR}) it follows that  
the shadowing correction to nucleonic $\Delta xF_3$ extracted 
from nuclear data is
\beq
\delta(\Delta xF_3)={Q^2\over 4\pi^2\alpha_W}{1\over A}
\left(\delta \sigma^A_{L}-\delta \sigma^A_{R} \right).
\label{eq:DELF3}
\eeq 
 Obviously, the  shadowing correction to $\Delta x F_3$
 is  related to $\delta x F_3$,
\beq
\delta x F_3={1\over 2}\delta(\Delta x F_3).
\label{eq:DELDEL}
\eeq 
To give an idea of the magnitude of the shadowing
 effect we evaluate the ratio of the nuclear shadowing correction,
$\delta(\Delta x F_3)$,
 to the 
nuclear structure function of the impulse approximation, $A\Delta x F_3$,
\beq
R={\delta(\Delta x F_3)\over A\Delta x F_3}=
{\delta\sigma^A_L-\delta\sigma^A_R\over
A\sigma_L-A\sigma_R}.
\label{eq:R}
\eeq
We calculate $R$ as a function of $Q^2$ for several values 
of Bjorken $x$ in the 
kinematical range of CCFR/NuTeV experiment. 
Our results obtained for  realistic nuclear densities of 
Ref.\cite{DEVRIES} are presented 
 in Figure \ref{fig:RNUC}.  Shown is  
the ratio $R(Q^2)$ for different nuclear 
targets including $^{56}Fe$.

We evaluate the ratio $R$ making use of
the color dipole factorization as described above. The differential
gluon density ${{\cal F}(x_g,\kappa^2)}$ in Eq.~(\ref{eq:SIGMA})  
was determined in  Ref.~\cite{NIKIV}. 
It is a function
of the gluon momentum fraction, $x_g$. For our purposes it suffices 
to use the approximation 
\beq
x_g=2x\left(1+{M^2\over 2Q^2}\right),
\label{eq:XG} 
\eeq
where $M^2=(m + \mu)^2$. The constituent $u$-,$d$-,$s$- and $c$-quark 
masses we use  are  $0.2$, $0.2$, $0.35$ and  $1.3$ GeV, 
respectively \cite{FZ}. 
  At $Q^2\gg M^2$ Eq.~(\ref{eq:XG}) reduces to $x_g=2x$
corresponding to the collinear DLLA. 
The longitudinal momentum transfer entering the nuclear  
form factor ${\cal S}_A$ is
\beq
k_L=x_gm_N.
\label{eq:KAPLX}
\eeq
One more simplification is that instead of the numerical Fourier transform
of the realistic nuclear density, for ${\cal S}_A$  we take a Gaussian 
parameterization
normalized to the correct nuclear charge radius $R_A$. 

At small $x$ and  high $Q^2$  the shadowing correction
scales, $\delta\sigma_{L,R}\propto 1/Q^2$.
 The absorption cross section $\sigma_{L,R}$ 
scales as well. The ratio $\delta\sigma_{L,R}/\sigma_{L,R}$ 
slowly decreases with growing $Q^2$ because 
of the logarithmic scaling violation in $\sigma_{L,R}$.
Toward the region of $x>0.01$,
 both the nuclear form factor and the mass threshold effect
suppress $R$ at $Q^2\lsim M^2$ (see Fig.\ref{fig:RNUC}).

\section{Summary}

In this paper we have presented  the color dipole analysis of nuclear 
effects in charge current DIS. The emphasis was put on
the pronounced effect of  left-right  asymmetry of shadowing in
neutrino-nucleus DIS at small values of Bjorken $x$. The L-R asymmetry of
$\nu N$  interactions  is quantified
in terms of the conventional structure function $xF_3$. The latter
is dominated by the  diffractive excitation of highly asymmetric $c\bar s$
component of the light-cone $W^+$.
We predicted strikingly
different scaling behavior of nuclear shadowing for the left-handed and
right-handed $W^+$. Large, about $20-25\%$, shadowing in the Fe structure functions
is predicted, which is important for a precise determination of
the nucleon structure functions $xF_3$ and $\Delta xF_3$.

{\large\bf Appendix.}
The $W^+\to c\bar s$-transition vertex is 
$$gU_{cs}\bar c\gamma_{\mu}(1-\gamma_5)s, $$
where $U_{cs}$ is an element of the CKM-matrix and the weak charge $g$ is
related to the Fermi coupling constant $G_F$ through the equation 
\beq
{G_F\over \sqrt{2}}={g^2\over m^2_{W}}.
\label{eq:GF}
\eeq
 The polarization states of W-boson 
 carrying the laboratory frame four-momentum 
\beq
q=(\nu,0,0,\sqrt{\nu^2+Q^2}) 
\label{eq:Q}
\eeq
are described 
by the four-vectors $e_{\lambda}$, with
\bea
e_0={1\over Q}(\sqrt{\nu^2+Q^2},0,0,\nu), \nonumber\\
e_{\pm}=\mp{1\over \sqrt{2}}(0,1,\pm i,0), 
\label{eq:WPOL}
\eea
the unit vectors $\vec e_x$ and $\vec e_y$ being in $q_x$- and $q_y$-direction,
respectively. 
 Then, the vector $(V)$ and axial-vector $(A)$ components of the 
light-cone wave function 
\beq
\Psi_{\lambda}^{\lambda_1,\lambda_2}(z,{\bf r})=
V_{\lambda}^{\lambda_1,\lambda_2}(z,{\bf r})  -
A_{\lambda}^{\lambda_1,\lambda_2}(z,{\bf r}),
\label{eq:PSIVA}
\eeq
are as follows \cite{FZ}:
\bea
V_0^{\lambda_1,\lambda_2}(z,{\bf r})={\sqrt{\alpha_W N_c}\over 2\pi Q}
\left\{
\delta_{\lambda_1,-\lambda_2}\left[
2Q^2z(1-z)
\right.
\right.
 \nonumber\\
\left.
\left.
+(m-\mu)[(1-z)m-z\mu]\right]K_0(\varepsilon r)
\right.
 \nonumber\\
\left.
-i\delta_{\lambda_1,\lambda_2}(2\lambda_1)e^{-i2\lambda_1\phi}(m-\mu)
\varepsilon K_1(\varepsilon r)\right\}\,,
\label{eq:V0}
\eea
\bea
A_0^{\lambda_1,\lambda_2}(z,{\bf r})={\sqrt{\alpha_W N_c}\over 2\pi Q}
\left\{
\delta_{\lambda_1,-\lambda_2}(2\lambda_1)\left[
2Q^2z(1-z)
\right.
\right.
 \nonumber\\
\left.
\left.
+(m+\mu)[(1-z)m+z\mu]\right]K_0(\varepsilon r)
\right.
 \nonumber\\
\left.
+i\delta_{\lambda_1,\lambda_2}e^{-i2\lambda_1\phi}(m+\mu)
\varepsilon K_1(\varepsilon r)\right\}\,.
\label{eq:A0}
\eea
The Eqs.~(\ref{eq:V0},\ref{eq:A0}) describe
 scalar and  axial quark-antiquark excitations of $W^+$.
For  the  right- and left-handed  $W^+$ corresponding to  
$\lambda=\pm 1$ we obtain 
\bea
V_{\lambda}^{\lambda_1,\lambda_2}(z,{\bf r})=
-{\sqrt{2\alpha_W N_c}\over 2\pi }\left\{
\delta_{\lambda_1,\lambda_2}\delta_{\lambda,2\lambda_1}[(1-z)m+z\mu]
 K_0(\varepsilon r)
\right.
\nonumber\\
\left.
-i(2\lambda_1)\delta_{\lambda_1,-\lambda_2}e^{i\lambda\phi}
\left[(1-z)\delta_{\lambda,-2\lambda_1}+z\delta_{\lambda,2\lambda_1} \right]
\varepsilon K_1(\varepsilon r)\right\}
\label{eq:VRL}
\eea
and 
\bea
A_{\lambda}^{\lambda_1,\lambda_2}(z,{\bf r})=
{\sqrt{2\alpha_W N_c}\over 2\pi }\left\{
\delta_{\lambda_1,\lambda_2}\delta_{\lambda,2\lambda_1}(2\lambda_1)
[(1-z)m-z\mu]
 K_0(\varepsilon r)
\right.
\nonumber\\
\left.
+i\delta_{\lambda_1,-\lambda_2}e^{i\lambda\phi}
\left[(1-z)\delta_{\lambda,-2\lambda_1}+z\delta_{\lambda,2\lambda_1} \right]
\varepsilon K_1(\varepsilon r)\right\}\,,
\label{eq:ARL}
\eea
where
\beq
\varepsilon^2=z(1-z)Q^2+(1-z)m^2+z\mu^2
\label{eq:VAREPS}
\eeq
 and   $K_{\nu}(x)$ is the modified Bessel function. 
We  consider only  Cabibbo-favored transitions and
$$\alpha_W={g^2/4\pi}.$$ 
The quark 
and anti-quark masses are $m$ and $\mu$, respectively. The 
 azimuthal angle of ${\bf r}$ is denoted by  $\phi$. 
To switch $W^+\to W^{-}$ one should perform the replacement 
$m\leftrightarrow\mu$ in the equations above.
The light-cone description of the neutral current (NC) interactions
mediated by the $Z$-boson transition, $Z\to q{\bar q}$,
  is an obvious extension of
 Eqs.~(\ref{eq:V0},\ref{eq:A0},\ref{eq:VRL},\ref{eq:ARL}), where
one should equate quark masses, $m=\mu$, and multiply vector and axial-vector
components of $\Psi_{\lambda}^{\lambda_1,\lambda_2}$ by corresponding 
NC coupling constants.

\end{document}